# The Role of AI in Facilitating Interdisciplinary Collaboration: Evidence from AlphaFold

Naixuan Zhao[1], Chunli Wei[1,2], Xinyan Zhang[3] and Jiang Li [1,*]

[1] School of Information Management, Nanjing University, Nanjing, China

[2] Department of Information Systems, City University of Hong Kong, Hong Kong, China

[3] Faculty of Business - Marketing, Hong Kong Polytechnic University, Hong Kong, China

* Corresponding author: lijiang@nju.edu.cn

The acceleration of artificial intelligence (AI) in science is recognized and many scholars have begun to explore its role in interdisciplinary collaboration. However, the mechanisms and extent of this impact are still unclear. This study, using AlphaFold's impact on structural biologists, examines how AI technologies influence interdisciplinary collaborative patterns. By analyzing 1,247 AlphaFold-related papers and 7,700 authors from Scopus, we employ bibliometric analysis and causal inference to compare interdisciplinary collaboration between AlphaFold adopters and non-adopters. Contrary to the widespread belief that AI facilitates interdisciplinary collaboration, our findings show that AlphaFold increased structural biology-computer science collaborations by just 0.48%, with no measurable effect on other disciplines. Specifically, AI creates interdisciplinary collaboration demands with specific disciplines due to its technical characteristics, but this demand is weakened by technological democratization and other factors. These findings demonstrate that artificial intelligence (AI) alone has limited efficacy in bridging disciplinary divides or fostering meaningful interdisciplinary collaboration.

## 1 Introduction

The development of artificial intelligence (AI) has demonstrated immense potential across various fields, accelerating and even revolutionizing progress in different domains while also driving advancements in scientific research (Gao & Wang, 2024; Wang, 2024). Over an extended period, AI has primarily functioned as a data mining tool in scientific research, typically regarded as a supplementary analytical method. For instance, in the biomedical field, scientists have utilized AI techniques such as Convolutional Neural Networks (CNNs), Support Vector Machines (SVMs), and Genome-Wide Association Studies (GWAS) to tackle diverse tasks, like categorizing medical images and identifying links between genetic factors and other biological traits. However, AI has now evolved to play a more integral role in the production of scientific knowledge (Lin et al., 2025). It can now emulate human reasoning and even predict potential future scientific advancements (Sourati et al., 2023). The impact of AI on scientific progress is profound, recent research indicates that AI is fundamentally transforming the scientific process, influencing everything from defining research priorities and shaping hypotheses (Banker et al., 2024) to conducting experiments (Ziems et al., 2024), disseminating knowledge (Van & Perkel, 2023), and engaging the public (Floridi et al., 2018). These applications are reshaping scientific practices in significant ways.

As one of AI's most emblematic triumphs, AlphaFold has greatly promoted the development of structural biology and other related fields (Ren et al., 2023). AlphaFold was first introduced in the journal *Nature* in January 2020 (Senior et al., 2020), followed by the release of AlphaFold2 in July 2021 (Jumper et al., 2021). The principal developers of AlphaFold2 were subsequently awarded the 2024 Nobel Prize in Chemistry for their work. The original AlphaFold primarily relied on the Transformer architecture and graph neural networks, significantly improving the accuracy and efficiency of protein structure prediction (Varadi & Velankar, 2023). In May 2024, the DeepMind team unveiled AlphaFold3, which is based on a diffusion model, along with a web-based tool (Abramson et al., 2024). Nevertheless, for those structural biologists working on protein prediction, leveraging AlphaFold for research remains challenging. The computational demands of AlphaFold2 pose a significant barrier, as AlphaFold2 running the model relies on high-end GPUs with substantial memory capacity, which discourages the attempts of researchers in small labs or underfunded institutions in using AlphaFold (Le et al., 2025). Although AlphaFold3 offers a web-based tool to reduce accessibility costs, those structural biologists still need a basic understanding of machine learning workflows to use it effectively (Fleming et al., 2025). Therefore, for structural biologists who wish to utilize AlphaFold but lack the necessary technical resources or expertise, it is supposed that collaborating with computer science experts becomes a viable option. This collaborative demand, driven by the inherent characteristics of AI tools, elevates AlphaFold beyond its function as a mere prediction tool, transforming it into a nexus connecting different disciplines and stimulating momentum for interdisciplinary collaboration.

Although research on the impact of AlphaFold on interdisciplinary collaboration is still relatively scarce in academia. However, it is evident that the development of AlphaFold requires the joint efforts of scholars from the fields of biology and computer science (Tunyasuvunakool et al., 2021; Jumper et al., 2021). From the joint curation of training corpora in protein databases, through algorithmic innovations in model architecture, to experimental validation by structural biologists, every stage of AlphaFold traverses computational and life sciences (Varadi & Velankar, 2023; Gong et al., 2025). Its very success intimates that AI tools carry an intrinsic disposition for interdisciplinary collaboration. Echoing this, prevailing scholarly accounts contend that AI increasingly fosters interdisciplinary integration (Bahroun et al., 2023; Park & Choi, 2024). Extending this line of reasoning, curiosity arises as to whether AI operates as an inherent catalyst for such collaboration. Against this backdrop, this study takes AlphaFold's impact on structural biology as its starting point to explore the research question: *Does and to what extent AI facilitate interdisciplinary collaboration?*

## 2 Literature review

### *2.1 Artificial intelligence and scientific collaboration*

With the rapid advancement of artificial intelligence (AI) technologies, their impact on human labour and innovation has become increasingly significant. Existing research has extensively discussed the role of AI in productive labour (Standaert & Andries, 2026; Xing et al., 2026), generally distinguishing between two mechanisms: substitution and augmentation (Raisch & Krakowski, 2021). Based on this perspective, the study by Fügener et al. (2025) have found that in contexts characterized by high inter-task complementarity, AI tends to exert a stronger substitutive effect, whereas in contexts with high intra-task complementarity, AI more often plays an augmentative role. However, in the domain of scientific research, a distinctive form of knowledge work, the mechanisms through which AI operates remain highly uncertain. Unlike production labour at the enterprise level, scientific research is not only a cognitively intensive activity but also a deeply social and collaborative process (Hackett, 2005).

The extant literature generally suggests that AI in scientific research exhibits features of augmentation rather than substitution. Specifically, research activities can be conceptualized as a system composed of several interrelated sub-tasks, including problem formulation, hypothesis development, data collection and analysis, experimental validation, and manuscript writing (Moring, 2014). Within these processes, AI can provide assistance or even partial substitution in specific tasks such as data annotation, pattern recognition, code generation, and text refinement (Lehr et al., 2024; Biyela et al., 2024; Ismail et al., 2025). Nevertheless, the overall logic and creative nature of scientific inquiry make it difficult for AI to fully replace human researchers. For instance, although AI can efficiently organize literature, analyse data, and generate text, its outputs still require human validation, integration, and interpretation (Pedota et al., 2025). Therefore, within the "substitution-augmentation" framework, the role of AI in scientific research is primarily augmentative rather than substitutive. From this perspective, one may infer that AI is unlikely to diminish the necessity or frequency of scientific collaboration, as it does not replace the cognitive labour or social interactions of scientists. On the contrary, by improving efficiency and reducing collaboration costs, AI may indirectly facilitate greater levels of scientific cooperation. Existing empirical research also supports this inference. Ding et al. (2025) found that the application of generative AI not only failed to weaken scientific collaboration but actually contributed to the expansion of team size and the enhancement of international cooperation. This suggests that, as an augmentative technology, AI may promote the expansion of collaborative research networks by lowering the costs of communication and knowledge integration.

Overall, there remains a lack of systematic empirical research on its effects on scientific collaboration, although the application of AI in scientific activities is becoming increasingly widespread. Accordingly, one of the theoretical contributions of this study lies in providing new perspectives and empirical evidence on how AI influences scientific collaboration, particularly interdisciplinary collaboration, and in clarifying whether AI in scientific research functions primarily as a substitute or as an enhancer of human creativity through the lens of the "substitution-augmentation" framework.

*2.2 Impact of technology on interdisciplinary collaboration*

As scientific and societal challenges grow increasingly complex, interdisciplinary collaboration has become essential—reshaping the way knowledge is produced and problems are solved. Interdisciplinary collaboration hence becomes a key approach to solving these problems (Benoliel & Somech, 2015). It is explained by Jones (2009)'s "burden of knowledge": as the accumulation of knowledge increases, the amount of knowledge that new researchers entering a particular field need to master also increases. This makes it difficult for a single discipline's knowledge system to effectively address complex problems. A substantial body of research has already demonstrated that interdisciplinary collaboration brings numerous benefits to research, by integrating knowledge (Brown et al., 2023), optimizing resources (Van, 2015), solving comprehensive problems (Barabas et al., 2021), enhancing creativity (Mahringer et al., 2023), etc. Relevant scientometric studies of have also confirmed that interdisciplinary collaboration can indeed lead to more impactful research (Nissani, 1997, Specht & Crowston, 2022).

Some reasons are identified for researchers to engage in interdisciplinary collaboration: (1) the motivation of individual researchers, (2) the need to address complex problems, and (3) the advancement of academic fields (Vladova et al., 2025). Mechanisms are proposed to explain interdisciplinary collaboration: the theory-method interdisciplinary collaboration mode and the technology interdisciplinary collaboration mode (Dai and Boos, 2019). The former focuses on the integration of theories and methods from different disciplines, using interdisciplinary theoretical frameworks and research methods to solve complex problems. The latter, on the other hand, focuses on the combination of technologies from different disciplines, using interdisciplinary technological means to achieve innovation and breakthroughs. These mechanisms provide different pathways for interdisciplinary collaboration, facilitating the addressing of diverse scientific and societal challenges (Van & Hessels, 2011).

As one of the core driving forces of modern scientific research, technology has a particularly significant impact on interdisciplinary collaboration, and many scholars have conducted in-depth explorations on this topic. From this perspective, the most essential demand for interdisciplinary collaboration stems from the limitations of single-discipline technologies (Xiao et al., 2025). When research in a particular discipline encounters a bottleneck or cannot solve complex problems through existing technologies within that discipline, the introduction of technologies from other disciplines becomes an inevitable choice (Li et al., 2024). For example, although the development of quantum computing technology has brought great potential to the field of computer science, its practical application requires the joint collaboration of physicists, computer scientists, and materials scientists to overcome technical challenges and achieve commercial application (Gyongyosi & Imre, 2019). As a cross-cutting general-purpose technology, AI has likewise dismantled barriers between disciplines, promoting communication and collaboration across different fields (Xu et al., 2024). During the COVID-19 pandemic, medical researchers actively sought collaboration with AI experts to predict the spatiotemporal dynamics of disease transmission (Abbonato et al., 2024), thereby addressing challenges that were difficult to overcome within the medical field alone.

In summary, technology promotes interdisciplinary collaboration because the limitations of technology within a single discipline create the need to integrate knowledge and methods from different fields. However, researchers often find it difficult to possess expertise across multiple domains, which in turn motivates them to engage in collaborative innovation to address complex problems.

*2.3 AlphaFold and interdisciplinary collaboration*

Against the backdrop of the rapid development of artificial intelligence technologies, DeepMind's launch of AlphaFold in 2020 marked a major breakthrough in the field of structural biology. Prior to the emergence of AlphaFold, protein structure prediction typically required structural biologists to identify functional and stable regions of proteins, collect protein sequence and experimental structure information, perform modeling, and analyze structural data to interpret the biological functions of proteins (Yu, 2025). This process often took months or even years and relied on expensive experimental equipment such as X-ray crystallography and cryo-electron microscopy (Yu, 2025; Fersht, 2021). However, with the advent of AlphaFold, the accuracy and speed of protein structure prediction have been significantly improved, and its predictive results have, in some cases, reached accuracies comparable to those obtained through manual experiments (Miao et al., 2023; Varadi & Velankar, 2023; Szczepski & Jaremko, 2025). This has greatly enhanced research efficiency and lowered the technical barriers to participation in protein structure studies (Fersht, 2021; Szczepski & Jaremko, 2025; Thangamalai et al., 2025;).

Nevertheless, the predictive results of AlphaFold still need to be verified and refined through experimental means by structural biologists (Varadi & Velankar, 2023). The biological interpretation and functional attribution of its predicted structures continue to rely on expert human judgment. Moreover, although AlphaFold performs exceptionally well in predicting rigid and globular protein structures, its accuracy decreases significantly when dealing with flexible proteins, membrane proteins, and multi-domain complex proteins (Robson, 2022; Szczepski & Jaremko, 2025). The training of the AlphaFold model also depends on existing protein structure databases (Bertoline et al., 2023; Fadahunsi et al., 2024; Uhlmann et al., 2024); therefore, for novel structural domains that have not previously appeared, its predictive performance is relatively limited, and structural biologists are still required to provide training data (Robson, 2022). Taken together, it is evident that AlphaFold has not completely replaced the professional expertise and experimental capabilities of structural biologists, but rather serves an augmentative role in specific stages of research. This aligns with the "substitution-augmentation" framework discussed earlier (Raisch & Krakowski, 2021), suggesting that AlphaFold has not replaced researchers, scientific labor but has enhanced human-machine collaboration in research by improving efficiency and precision. Therefore, AlphaFold does not replace the work of structural biologists and thus does not weaken collaborative relationships among researchers.

At the same time, the operation of AlphaFold requires substantial computational power, and as the length of protein sequences increases, the structure prediction process becomes extremely time-consuming (Bertoline et al., 2023). Although AlphaFold3 has partially lowered the threshold for use through a web-based interface, researchers still need to possess basic knowledge of machine learning workflows and data processing to use the tool effectively (Fleming et al., 2025). Many researchers, especially experimental scientists who have not received training in bioinformatics, have encountered huge obstacles when using these tools to conduct their own research (Elfmann & Stülke, 2025). This means that for structural biologists lacking a background in computer science or artificial intelligence, the application of AlphaFold in fact creates a demand for interdisciplinary collaboration. They often need to work with computer scientists, algorithm engineers, or data scientists to optimize model performance, improve prediction workflows, and interpret model results. This trend supports the earlier argument that technology promotes interdisciplinary collaboration, that when the technologies of a single discipline cannot fully meet research needs, collaboration across different fields becomes a key driver of scientific progress.

In summary, as a typical augmentative artificial intelligence technology, AlphaFold has not replaced the scientific labor of structural biologists but has instead fostered the need for interdisciplinary collaboration with computer science. Therefore, the following hypothesis is proposed: *the application of AlphaFold promotes interdisciplinary collaboration, particularly between structural biologists and computer scientists*.

## 3 Data and methods

### *3.1 Author data collection*

To identify authors who utilized AlphaFold in their research, we first need to retrieve relevant studies employing AlphaFold. Therefore, we conducted a search on Scopus with the query: TITLE-ABS-KEY('alphafold') on November 13, 2024, which yielded 1,500 publications. Using Scopus' export function, we retrieved all available metadata for these records (including author IDs, publication years, institutional affiliations, and journal information). To ensure data accuracy for subsequent analysis, we manually accessed the full-text links of each publication to review the full texts of these papers and performed a manual screening to exclude irrelevant studies (e.g., papers that mentioned AlphaFold only in the title, abstract, or keywords but did not actually use it in their research). This refinement resulted in 1,247 relevant papers. Based on the author

IDs and corresponding contact information of these papers, we compiled a dataset of 7,700 authors, including 1,371 corresponding authors.

The RCSB PDB database (https://www.rcsb.org/) is a repository where structural biologists deposit experimentally determined or computationally predicted protein structures. As of now, the database contains 232,095 protein structures. Besides recording the data of protein structure, RCSB PDB also documents the associated research papers, which we define as "PDB papers." Using the DOI and PubMed ID of these papers, we retrieved 85,169 publications from Scopus. Subsequently, we extracted the corresponding authors of each paper and, after removing duplicates, obtained 19,383 unique corresponding author IDs, representing 19,383 principal investigators (PIs) in structural biology. This dataset serves as the basis for our comparative analysis.

### *3.2 Construction of treatment and control groups*

We designed a quasi-experiment and constructed treatment and control groups following the methodology of Yu (2025). The criteria for defining the treatment and control groups were as follows: We selected 2020 as the cutoff year (since AlphaFold was officially released in January 2020) and required that both groups had published protein structure-related papers (PDB papers) both before and after 2020. The key distinction was that treatment-group PIs had to have published AlphaFold-related papers after 2020, whereas control-group PIs had no such publications. Applying these criteria to the 19,383 structural biologists, we ultimately identified: 692 PIs in the treatment group (those who adopted AlphaFold after 2020), and 2,822 PIs in the control group (those who did not use AlphaFold after 2020). This stratification allows for a comparative analysis of collaboration patterns between early AlphaFold adopters and non-adopters.

### *3.3 Determination of authors' disciplines*

After establishing the treatment and control groups, we need to determine the disciplines of these structural biologists' collaborators to identify whether interdisciplinary collaboration has occurred. We assessed collaborators' fields based on the disciplinary distribution of their historical publications. The specific classification rules are defined as follows: we categorize authors according to the proportional distribution of disciplines across all their historical publications. If biomedical papers account for more than 50% of an author's publications, he/she is classified as belonging to the biomedical field. Those with a more dispersed publication distribution where no single field exceeds 50% are classified as belonging to other fields. Additionally, we evaluated computer science knowledge for subsequent descriptive statistics and causal inference. The assessment criteria were defined as follows: if an author has published two or more papers in computer science, they are considered to possess computer science knowledge.

In the actual implementation process, we first obtained all collaborator IDs from both the treatment and control groups, then retrieved and exported their complete publication records from Scopus using these IDs. However, we discovered that Scopus does not provide disciplinary classification information for these publications, necessitating manual categorization. To address this, we employed a machine learning approach for disciplinary classification of papers. Building upon the research of Lv et al. (2022), we implemented the SCIBERT algorithm (Beltagy et al., 2019), and utilized their provided dataset for training the classification model. This dataset comprises 441,216 papers published in the Web of Science Core Collection in 2016, with each paper assigned to a single disciplinary category (single-label classification) through the ECOOM classification method (Glänzel & Schubert, 2003). The SCIBERT algorithm achieved an overall accuracy of 81.57% when trained with an 80:20 split between training and test sets. The relatively lower accuracy stems partly from the dataset's more granular classification system for biomedical fields, where the algorithm's precision drops to approximately 70%, while maintaining higher accuracy (around 90%) in other disciplinary categories. Since our study does not require detailed sub-classification within biomedical fields, the SCIBERT-based approach proves sufficiently reliable for our purposes. We subsequently applied this classification system to categorize all collaborators' publications and determine their disciplinary affiliations. Additionally, we validated the disciplinary classifications for authors in both treatment and control groups, with results confirming their consistent classification within biomedical fields, thereby providing evidence supporting the accuracy of our classification methodology.

### *3.4 Regression model*

We employed a DID model to evaluate the impact of AlphaFold on interdisciplinary collaboration. We conducted two parallel regression analyses: the first examined whether AlphaFold adoption influenced structural biologists' propensity for general interdisciplinary collaboration (using *Interdis_proportion* as the dependent variable), while the second specifically assessed its effect on collaboration with computer scientists (using *Com_proportion* as the dependent variable). The key independent variable in both models were whether the researcher had published AlphaFold-related papers (*AF_experience*). Drawing on Wooley et al. (2015) and this study, we controlled for variables including the researcher's publication output (*Paper_count*) and academic age (*Academic_age*). Additionally, we controlled for whether the authors themselves had computer-science experience *(Com_experience*) and for the proportion of their collaborators who possessed computer experience but were not from computer-science (*Coll_experience*). Controlling for *Com_experience* helps account for the possibility that authors with computational expertise may have less need to collaborate with computer scientists, as they can handle computational aspects themselves (see Section 5.1 for more accurate discussion). *Coll_experience* was included because structural biologists may preferentially collaborate with colleagues who possess computational skills even outside the computer-science field, which could independently influence the dependent variable. All control variables were measured as time-varying covariates and the linear regression models are as follows,

$$Interdis_proportion_{it} = \alpha + \beta(Treat_i \times Post_t) + \gamma X_{it} + \mu_i + \lambda_t + \varepsilon_{it} \quad (1)$$

$$Com_proportion_{it} = \alpha + \beta(Treat_i \times Post_t) + \gamma X_{it} + \mu_i + \lambda_t + \varepsilon_{it} \quad (2)$$

where $Treat_i$ is a dummy variable indicating $AF_experience$, $Post_t$ marks the temporal cutoff (1=after 2020, 0=before 2020). $Treat_i \times Post_t$ is the interaction term capturing AlphaFold's adoption effect, $\alpha$ denotes the intercept term, and $\beta$ is the key coefficient of interest to be estimated. $X_{it}$ represents the set of control variables described above, which are sequentially added in the regression models. $\mu_i$ and $\lambda_t$ indicate individual (*Individual FE*) and time fixed effects (*Year FE*), respectively, and $\varepsilon_{it}$ represents the error term.

Table 1. Definition of the variables.

| Name | Type | Definition |
|---|---|---|
| **Dependent variables** | | |
| *Interdis _proportion* | Continuous | Proportion of collaborators from disciplines other than structural biology |
| *Com_proportion* | Continuous | Proportion of collaborators from computer science |
| **Independent variables** | | |
| *AF_experience* | Dummy | Whether the scientist have published AlphaFold related articles: 1 = yes, 0 = no |
| **Control variables** | | |
| *Com_experience* | Dummy | Whether the scientist has computer science experience: 1 = yes, 0 = no |
| *Paper_count* | Continuous | Publication counts of scientists (Wooley et al., 2015) |
| *Academic_age* | Continuous | Academic age of scientists (Wooley et al., 2015) |
| *Coll_experience* | Continuous | Proportion of collaborations with researchers who have computer skills but non-CS backgrounds |

## 4 Results

### *4.1 Descriptive statistics*

We first conducted a preliminary analysis of the 1,247 AlphaFold-related papers. Figure 1a presents the temporal trend of AlphaFold-related publications since January 2020, when the DeepMind team first officially published AlphaFold in *Nature* (Senior et al., 2020). The number of relevant papers began to increase significantly around September 2021, which aligns with the typical publication cycle for academic research. This observed pattern demonstrates a one-to-two-year lag between the tool's initial release and widespread adoption in research publications, reflecting the time required for researchers to implement new methodologies and complete subsequent studies.

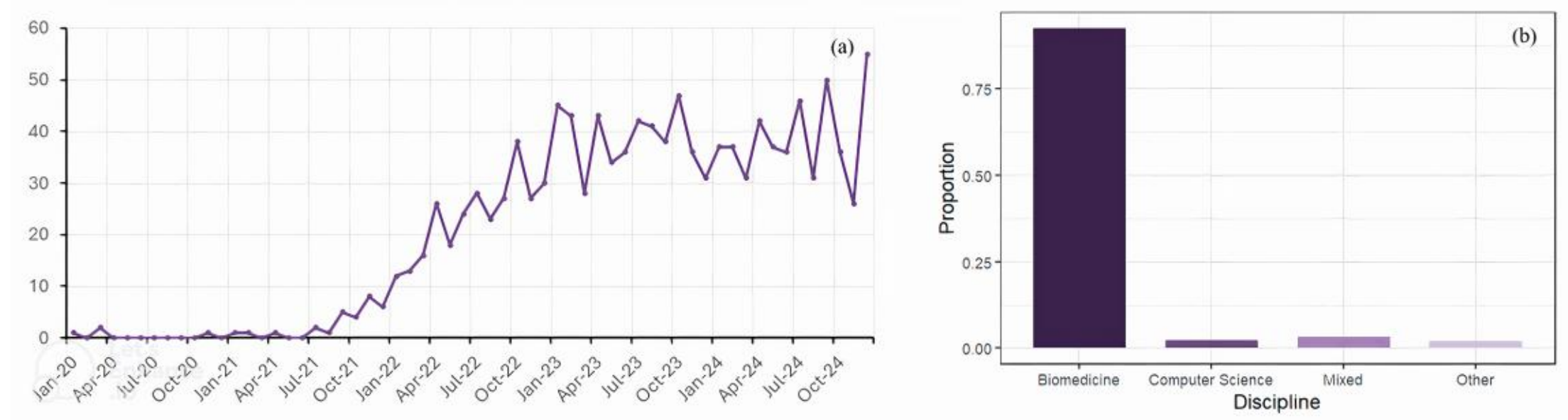

Figure 1. Publication and author distribution. (a) Temporal trend of AlphaFold papers. (b) Distribution of authors by discipline.

Table 2 presents the descriptive statistics of variables for both the treatment and control groups. Regarding the key variables, the average number of computer science collaborators in the treatment group post-2020 (0.0882) was significantly higher than that in the control group (0.0248), while no significant difference was observed in the proportion of collaborators from other disciplines between the two groups. This finding provides preliminary evidence that AlphaFold may promote collaboration with specific disciplines.

Table 2. Descriptive statistics.

| **Varaible** | **Group** | **Period** | **Obs** | **Mean** | **Std.Dev** | **Min** | **Max** |
|---|---|---|---|---|---|---|---|
| Number of Collaborators | experimental | Before 2020 | 692 | 9.4191 | 7.5316 | 1 | 53 |
| | | After 2020 | 692 | 10.5289 | 13.9188 | 0 | 159 |
| | control | Before 2020 | 2822 | 13.5429 | 15.5850 | 0 | 218 |
| | | After 2020 | 2822 | 14.5974 | 16.4283 | 0 | 177 |
| Number of Collaborators from other disciplines | experimental | Before 2020 | 692 | 0.1199 | 0.3630 | 0 | 3 |
| | | After 2020 | 692 | 0.4350 | 1.5860 | 0 | 15 |
| | control | Before 2020 | 2822 | 0.0092 | 0.1126 | 0 | 2 |
| | | After 2020 | 2822 | 0.5677 | 2.7500 | 0 | 52 |
| Number of Collaborators from Computer Science | experimental | Before 2020 | 692 | 0.0058 | 0.0759 | 0 | 1 |
| | | After 2020 | 692 | 0.0882 | 0.4463 | 0 | 5 |
| | control | Before 2020 | 2822 | 0.0092 | 0.1126 | 0 | 2 |
| | | After 2020 | 2822 | 0.0248 | 0.2099 | 0 | 6 |
| Number of Collaborators with Computer Experience | experimental | Before 2020 | 692 | 2.4783 | 2.7754 | 0 | 22 |
| | | After 2020 | 692 | 2.6329 | 7.7771 | 0 | 79 |
| | control | Before 2020 | 2822 | 2.5974 | 5.5359 | 0 | 129 |
| | | After 2020 | 2822 | 2.4111 | 4.9314 | 0 | 78 |
| *Paper_count* | experimental | / | 692 | 17.5043 | 23.4255 | 2 | 342 |
| | control | / | 2822 | 14.0578 | 15.4938 | 2 | 375 |
| *Academic_age* | experimental | / | 692 | 22.9176 | 12.1702 | 5 | 67 |
| | control | / | 2822 | 23.2612 | 10.2794 | 5 | 67 |
| *Com_experience* | experimental | / | 692 | 0.1113 | 0.3147 | 0 | 1 |
| | control | / | 2822 | 0.0559 | 0.2299 | 0 | 1 |

### *4.2 Regression results*

The regression results in Tables 3 and 4 reveal that the use of AlphaFold (*Treat×Post*) significantly increased the proportion of computer science collaborators among structural biologists by 0.48 percentage points ($p<0.001$), while its impact on the proportion of collaborators from other disciplines (*Interdis_proportion*) did not reach statistical significance ($p>0.05$). These results indicate that while AlphaFold has enhanced collaboration with computer science to some extent, it has not significantly improved overall interdisciplinary collaboration levels.

Table 3. Regression results of interdisciplinary collaboration with non-structural-biologists.

| | *Interdis_proportion* | | | | |
|---|---|---|---|---|---|
| | (1) | (2) | (3) | (4) | (5) |
| *Treat×Post* | 0.00492 | 0.00492 | 0.00492 | 0.00492 | 0.00491 |
| | (0.00345) | (0.00345) | (0.00345) | (0.00345) | (0.00345) |
| *Com_experience* | | 0.00521 | 0.00569 | 0.00577 | 0.00547 |
| | | (0.00277) | (0.00303) | (0.00303) | (0.00304) |
| *Paper_count* | | | -0.00002 | -0.00001 | -0.00000 |
| | | | (0.00004) | (0.00005) | (0.00005) |
| *Academic_age* | | | | -0.00003 | -0.00003 |
| | | | | (0.00008) | (0.00008) |
| *Coll_experience* | | | | | 0.00652 |
| | | | | | (0.00408) |
| *Individual FE* | Y | Y | Y | Y | Y |
| *Year FE* | Y | Y | Y | Y | Y |
| $R^2$ | 0.05009 | 0.05055 | 0.05058 | 0.05061 | 0.05094 |
| Observations | 7028 | 7028 | 7028 | 7028 | 7028 |

Notes. *** $p<0.001$, ** $p<0.01$, * $p<0.05$.

Table 4. Regression results of interdisciplinary collaboration with computer scientists.

| | *Com_proportion* | | | | |
|---|---|---|---|---|---|
| | (1) | (2) | (3) | (4) | (5) |
| *Treat×Post* | 0.00482*** | 0.00482*** | 0.00482*** | 0.00482*** | 0.00482*** |
| | (0.00073) | (0.00073) | (0.00073) | (0.00073) | (0.00073) |
| *Com_experience* | | 0.00090 | 0.00099 | 0.01040 | 0.00096 |
| | | (0.00059) | (0.00064) | (0.00064) | (0.00064) |
| *Paper_count* | | | -0.00000 | 0.00000 | 0.00000 |
| | | | (0.00000) | (0.00001) | (0.00001) |
| *Academic_age* | | | | -0.00002 | -0.00002 |
| | | | | (0.00001) | (0.00001) |
| *Coll_experience* | | | | | 0.00175* |
| | | | | | (0.00087) |
| *Individual FE* | Y | Y | Y | Y | Y |
| *Year FE* | Y | Y | Y | Y | Y |
| $R^2$ | 0.01807 | 0.01840 | 0.01842 | 0.01866 | 0.01924 |
| Observations | 7028 | 7028 | 7028 | 7028 | 7028 |

Notes. *** $p$<0.001, ** $p$<0.01, * $p$<0.05.

We further investigated the underlying reasons for this phenomenon. As a computationally intensive tool, AlphaFold's deployment and optimization critically depend on expertise in computer science, compelling structural biologists to collaborate with computer specialists. However, its technical requirements from other disciplines remain relatively low, failing to generate similar collaborative incentives. Secondly, as noted in Section 3.1, researchers with computer science backgrounds are significantly more likely to collaborate with computer scientists (45.10% vs. 16.86%). This homophily preference means AlphaFold primarily reinforces existing computer science collaboration networks rather than expanding interdisciplinary connections to other fields - a pattern further corroborated by the positive coefficients of the control variable *Com_experience* in Models 2-5 (Table 4).

On the other hand, while AlphaFold has facilitated collaboration between structural biologists and computer scientists, its actual impact remains relatively modest, increasing the proportion of computer science collaborators by only 0.48 percentage points. This limited effect can be attributed to several factors. First, structural biologists may prefer to learn and utilize AlphaFold through self-study, and the emergence of various alternative self-learning solutions (Fleming et al., 2025) has reduced the necessity for formal collaborations, thereby constraining further expansion of collaborative networks. Second, the development of simplified tools such as OpenFold and more user-friendly iterations like the web-based AlphaFold 3 has significantly lowered the technical barrier to entry (Le et al., 2025). As a result, structural biologists can now conduct relevant research independently without requiring cross-disciplinary collaboration.

### *4.3 Event Study*

To examine how AlphaFold influenced structural biologists' propensity to collaborate with computer scientists, we conducted an event study analysis. Figure 2 presents our findings, using 2019 (one year prior to AlphaFold's release) as the baseline year. The results largely satisfy the parallel trends assumption, supporting the validity of our empirical strategy. While the treatment effect remains statistically insignificant in 2020, we attribute this to the inherent publication delays in academic research (our subsequent analysis also confirms this). Additionally, it can be observed that the estimated coefficients show a declining trend from 2022 to 2024. We believe that this declining trend corresponds with the phased introduction of simplified tools like OpenFold, enhanced tutorials, and the web-based AlphaFold 3, all of which collectively lowered the technical barriers to adoption. As these user-friendly alternatives gained traction, structural biologists increasingly viewed computer science collaboration as non-essential rather than obligatory.

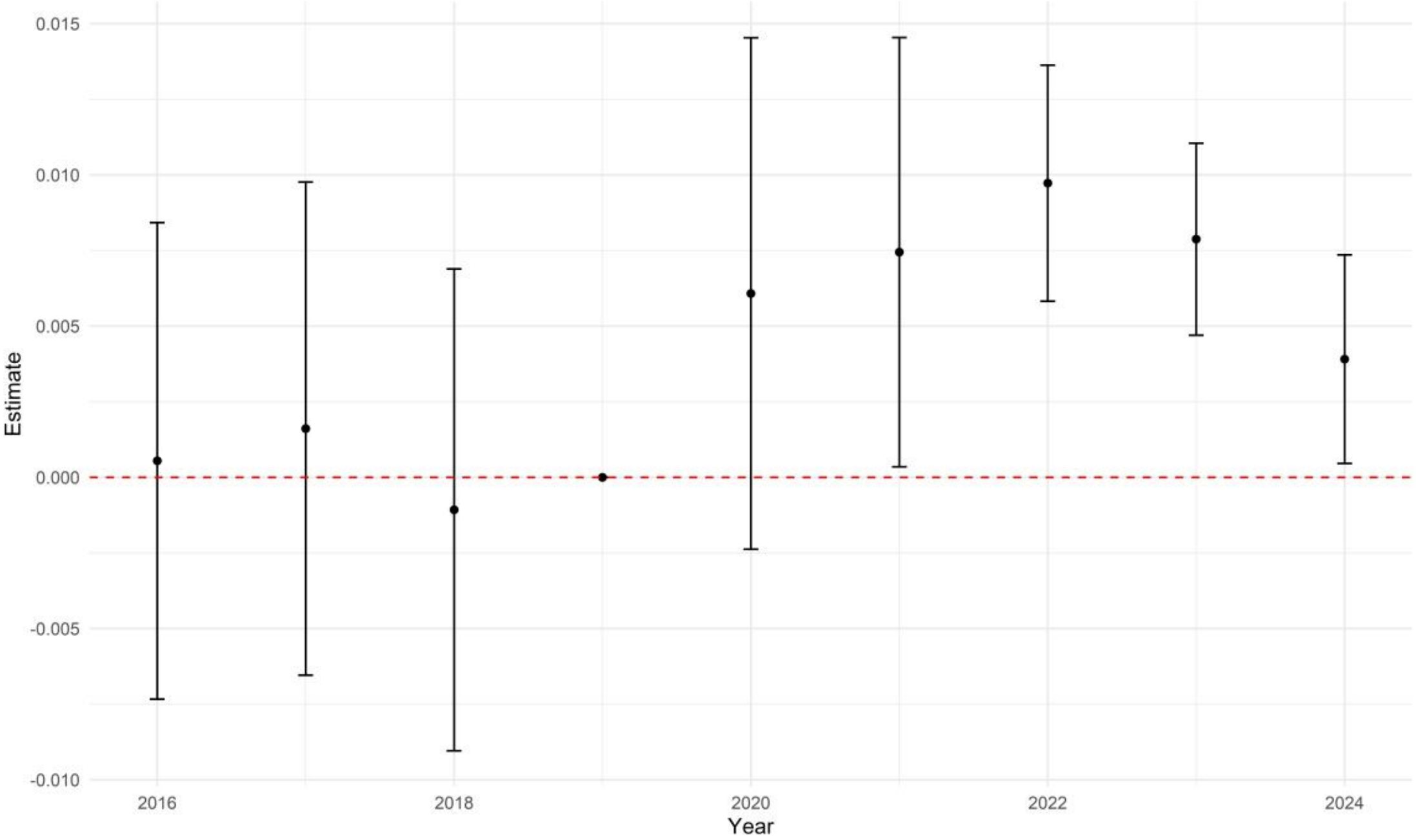

Figure 2. Event Study of AlphaFold's impact on structural biologists.

*4.4 Robustness check*

To ensure the rigor of our conclusions, we conducted additional robustness checks. The approach involved replacing the ECOOM classification system used in Section 3.3 to determine researchers' disciplinary affiliations with the disciplinary classification system provided by Web of Science. This classification system comprises approximately 250 disciplinary categories, and its classification of papers by discipline is essentially based on the journals to which the papers belong (Milojević, 2020). In other words, if a journal falls under the category of "Information Science & Library Science" all papers published in that journal will likewise be classified as "Information Science & Library Science" in the disciplinary field provided by Web of Science.

However, this classification criterion introduces two issues. First, due to its overly granular disciplinary classification standards, certain categories may suffer from insufficient training samples when using the Web of Science dataset for training, leading to inaccuracies in the classification results. Second, some journals are classified as belonging to multiple disciplines, but the papers published in them may only belong to one of those disciplines. Consequently, such papers may be incorrectly assigned to training sets for other disciplines, thereby affecting the classification results (Abramo et al.,2018; Shu et al., 2019). Our current solution was to exclude papers from multidisciplinary journals when training the SCIBERT model, using only those from single-discipline journals for training, even though this may result in some sample loss. These limitations are also why the main regression analysis in this paper did not adopt the Web of Science classification system but instead used it solely for robustness checks.

Table 5 and 6 presents the results of our robustness checks. As shown by the coefficient of the *Treat×Post* term in the table, our findings remain statistically significant ($p<0.001$) under the Web of Science classification scheme. Specifically, AlphaFold leads to an approximately 0.53 percentage point increase in the proportion of computer science collaborators among structural biologists, indicating that it has facilitated some degree of interdisciplinary collaboration between structural biology and computer science. This effect size is very close to the main regression result (0.482 percentage points), further demonstrating the robustness of our findings.

Table 5. Robustness check of interdisciplinary collaboration with non-structural-biologists.

| | | | *Interdis_proportion* | | |
|---|---|---|---|---|---|
| | (1) | (2) | (3) | (4) | (5) |
| *Treat×Post* | 0.00622 | 0.00622 | 0.00622 | 0.00622 | 0.00621 |
| | (0.00170) | (0.00170) | (0.00170) | (0.00170) | (0.00170) |
| *Com_experience* | | 0.00320 | 0.00320 | 0.00320 | 0.00320 |
| | | (0.00145) | (0.00146) | (0.00146) | (0.00146) |
| *Paper_count* | | | 0.00002 | 0.00002 | 0.00002 |
| | | | (0.00000) | (0.00000) | (0.00000) |
| *Academic_age* | | | | -0.00001 | -0.00001 |
| | | | | (0.00008) | (0.00008) |
| *Coll_experience* | | | | | 0.00305 |
| | | | | | (0.00408) |
| *Individual FE* | Y | Y | Y | Y | Y |
| *Year FE* | Y | Y | Y | Y | Y |
| $R^2$ | 0.05655 | 0.05671 | 0.05684 | 0.05697 | 0.05911 |
| Observations | 7028 | 7028 | 7028 | 7028 | 7028 |

Notes. *** $p<0.001$, ** $p<0.01$, * $p<0.05$.

Table 6. Robustness check of interdisciplinary collaboration with computer scientists.

| | | | *Com_proportion* | | |
|---|---|---|---|---|---|
| | (1) | (2) | (3) | (4) | (5) |
| *Treat×Post* | 0.00540*** | 0.00540*** | 0.00540*** | 0.00540*** | 0.00532*** |
| | (0.00036) | (0.00036) | (0.00036) | (0.00036) | (0.00036) |
| *Com_experience* | | 0.00125 | 0.00111 | 0.00111 | 0.00109 |
| | | (0.00029) | (0.00031) | (0.00031) | (0.00031) |
| *Paper_count* | | | 0.00000 | 0.00000 | 0.00000 |
| | | | (0.00000) | (0.00001) | (0.00001) |
| *Academic_age* | | | | 0.00000<br>(0.00001) | 0.00000<br>(0.00001) |
| *Coll_experience* | | | | | 0.00172**<br>(0.00585) |
| *Individual FE* | Y | Y | Y | Y | Y |
| *Year FE* | Y | Y | Y | Y | Y |
| $R^2$ | 0.02014 | 0.02032 | 0.02056 | 0.02077 | 0.02081 |
| Observations | 7028 | 7028 | 7028 | 7028 | 7028 |

Notes. *** *p*<0.001, ** *p*<0.01, * *p*<0.05.

### *4.5 Heterogeneity*

***Different productivity groups***

To examine how scholars with different productivity levels vary in their preferences for interdisciplinary collaboration, we conducted a heterogeneity analysis by dividing authors into high- and low-productivity groups and performing separate regression analyses for each group. Given that our analysis focuses on corresponding authors, who generally exhibit relatively high publication output, we classified authors with an average annual output of one or more papers as high-productivity researchers, while those averaging fewer than one paper per year were categorized as low-productivity researchers. Table 7 presents the regression results.

Table 7. Heterogeneity between different productivity groups.

| | *Com_proportion* | |
|---|---|---|
| *Productivity* | **High-productive**<br>(1) | **Low-productive**<br>(2) |
| *Treat×Post* | 0.00577***<br>(0.00164) | 0.00455***<br>(0.00082) |
| *Com_experience* | 0.00193<br>(0.00105) | 0.00027<br>(0.00081) |
| *Paper_count* | -0.00000<br>(0.00001) | 0.00001<br>(0.00003) |
| *Academic_age* | -0.00000<br>(0.00004) | -0.00003<br>(0.00002) |
| *Coll_experience* | 0.00335<br>(0.00204) | 0.00141<br>(0.00095) |
| *Individual FE* | Y | Y |
| *Year FE* | Y | Y |
| $R^2$ | 0.04816 | 0.01565 |
| Observations | 926 | 6102 |

Notes. *** *p*<0.001, ** *p*<0.01, * *p*<0.05.

The results reveal that the coefficient for high-productivity authors (0.00577) is slightly larger than that for low-productivity authors (0.00455), suggesting that when using

AlphaFold in their research, high-productivity authors exhibit a stronger preference for collaborating with computer science researchers compared to their low-productivity counterparts. This phenomenon may be primarily attributed to two factors: First, high-productivity authors typically possess more extensive professional networks and collaboration opportunities. Their established academic connections facilitate easier access to potential collaborators across different disciplines (Dietz & Bozeman, 2005). In contrast, low-productivity authors may have more limited networks, constraining their ability to identify and engage in interdisciplinary collaborations (Mitrović et al., 2023). Second, having demonstrated their research capabilities through sustained high output, high-productivity authors may be more inclined to explore novel research directions and collaboration models (Zeng et al., 2022). Low-productivity researchers, by comparison, might prefer to maintain stable research within familiar domains to avoid the uncertainties and risks associated with interdisciplinary collaboration.

***Different team size***

To further investigate how team size affects interdisciplinary collaboration, we conducted additional heterogeneity analysis at the team level. Adopting the classification criteria of Xu et al. (2024), we divided research teams into two groups based on the average number of authors per paper: large teams (with more than 3 authors on average) and small teams (with 3 or fewer authors on average). We then performed separate regression analyses for each group. Table 8 presents the regression results.

Table 8. Heterogeneity between different team sizes.

| | *Com_proportion* | |
|---|---|---|
| *Team Size* | **Large** ($X > 3$) | **Small** ($0 \leq X \leq 3$) |
| | (1) | (2) |
| *Treat×Post* | 0.00403*** | 0.00524*** |
| | (0.00107) | (0.00102) |
| *Com_experience* | -0.00072 | 0.00158 |
| | (0.00113) | (0.00081) |
| *Paper_count* | 0.00004 | 0.00000 |
| | (0.00004) | (0.00001) |
| *Academic_age* | -0.00001 | -0.00004 |
| | (0.00003) | (0.00002) |
| *Coll_experience* | 0.00078 | 0.00206 |
| | (0.00134) | (0.00115) |
| *Individual FE* | Y | Y |
| *Year FE* | Y | Y |
| $R^2$ | 0.01359 | 0.02406 |
| Observations | 3292 | 3736 |

Notes. *** *p*<0.001, ** *p*<0.01, * *p*<0.05.

The results show that the coefficient for small teams (0.00524) is higher than that for large teams (0.00403), indicating that when utilizing AlphaFold in research, small teams demonstrate a stronger propensity to seek collaboration with computer science scholars compared to large teams. This phenomenon may be attributed to the fundamental difference in resource allocation between team sizes: Small teams typically operate with limited resources and often lack specialized knowledge and technical support. This resource constraint motivates them to actively seek computer science collaborations to access necessary expertise. In contrast, large teams benefit from their abundant resources and thus reduce their reliance on external computer science partnerships when using AlphaFold.

***Different age groups***

Finally, we conducted a heterogeneity analysis of scholars across different academic age groups. The academic age groups were divided according to the quartiles of academic age, i.e., 5-17, 17-23, 23-30, and 30-67. The results of the heterogeneity analysis are presented in Table 9. From the table, it can be observed that scientists at different academic age levels show a trend of first increasing and then decreasing in their tendency for interdisciplinary collaboration. This phenomenon is similar to the research results of Van & Hessels (2011), which suggests that the relationship between interdisciplinary collaboration and academic age follows an inverted U-shaped curve, first increasing and then decreasing.

Table 9. Heterogeneity between different age groups.

| | *Com_proportion* | | | |
|---|---|---|---|---|
| *Academic Age* | **[0, 0.25)** ($5 \leq X < 17$) (1) | **[0.25, 0.5)** ($17 \leq X < 23$) (2) | **[0.5,0.75)** ($23 \leq X < 30$) (3) | **[0.75, 1]** ($30 \leq X \leq 67$) (4) |
| *Treat×Post* | 0.00528*** | 0.00731*** | 0.00383** | 0.00398*** |
| | (0.00158) | (0.00221) | (0.00126) | (0.00098) |
| *Com_experience* | -0.00030 | -0.00234 | 0.00107 | 0.00186** |
| | (0.00243) | (0.00236) | (0.00119) | (0.00058) |
| *Paper_count* | 0.00003 | 0.00001 | -0.00001 | 0.00001 |
| | (0.00007) | (0.00007) | (0.00002) | (0.00001) |
| *Academic_age* | 0.00011 | -0.00026 | -0.00005 | -0.00003 |
| | (0.00008) | (0.00039) | (0.00011) | (0.00003) |
| *Coll_experience* | 0.00318 | 0.00258 | -0.00014 | 0.00135 |
| | (0.00193) | (0.00248) | (0.00147) | (0.00114) |
| *Individual FE* | Y | Y | Y | Y |
| *Year FE* | Y | Y | Y | Y |
| $R^2$ | 0.02045 | 0.02855 | 0.01518 | 0.03238 |
| Observations | 1800 | 1702 | 1728 | 1798 |

Notes. *** $p$<0.001, ** $p$<0.01, * $p$<0.05.

The relatively low tendency for interdisciplinary collaboration among early-career scientists is mainly due to two aspects. On the one hand, scientists in the early stages of their academic careers have not yet accumulated sufficient resources and social connections, which to some extent limits their ability to engage in interdisciplinary collaboration (Bozeman et al., 2001; Bozeman & Corely, 2004). On the other hand, for scientists who are focused on a specific field, interdisciplinary research carries certain risks. Early-career scientists usually need more stable results to help them get promoted and pass evaluations, so they may be more inclined to focus on their familiar fields to ensure they can achieve reliable results (McLeish & Strang, 2016; Vladova et al., 2025). As scientists' academic age increases, they gradually accumulate rich social connections and resources, which provide a solid foundation for them to engage in interdisciplinary collaboration. Therefore, they are more willing to explore opportunities for interdisciplinary collaboration in their middle age to expand their research horizons and enhance their research impact (Abramo et al., 2014; Norton et al., 2023).

However, for older scientists, they have already accumulated enough resources and no longer need to rely on external help to carry out interdisciplinary collaboration. This is similar to the reason of team size, that is, with the accumulation of resources, scientists' needs and motivations for interdisciplinary collaboration may change. In addition, their research in the current discipline has become relatively mature and stable, and they may prefer to continue to delve into the research questions in their own field rather than explore interdisciplinary collaboration (Engwall, 2018; Nastar et al., 2018).

## 5 Discussion

### *5.1 Who are seeking interdisciplinary collaboration?*

AlphaFold-related papers primarily focus on the following topics: (1) direct application of AlphaFold for protein prediction in specific subfields; (2) algorithmic improvements to AlphaFold for particular subdomains; and (3) analysis and validation of protein structures predicted by AlphaFold. Among these, category (1) papers constitute the majority, accounting for approximately 70% (848 papers) of all studies. Notably, the first category (1) paper was published in September 2021, coinciding with the observed inflection point when AlphaFold-related publications began to exhibit significant growth.

We then performed descriptive statistics on authors who published AlphaFold-related papers. When scientists need to use new technologies like AlphaFold that have certain technical barriers but can significantly improve research efficiency, they usually adopt two strategies: either independently learning and exploring how to apply these new technologies, or seeking help from scientists with relevant technical backgrounds. When those who choose to seek help invite scholars from other disciplines to assist them, interdisciplinary collaboration emerges.

To observe interdisciplinary collaboration in AlphaFold-related papers, we applied the same method described in Section 2.3 to determine the disciplinary affiliations and assess computer science knowledge for all 7,700 authors of these papers. Among these 7,700 authors, 7,123 (92.5%) are from biomedical fields. Computer scientists account for a much smaller proportion, with only 182 scholars (2.4%). The numbers of authors from other disciplines or with mixed backgrounds are also relatively low, at 150 and 245 respectively (Figure 1b). Additionally, among all 7,700 authors, 22.2% have relevant computer science experience, meaning most authors do not possess advanced computer science knowledge.

Then, we investigated the collaborative patterns among these biomedical scholars. Among the 7,123 researchers from biomedical fields, 592 scholars chose to collaborate with computer scientists, demonstrating clear instances of interdisciplinary cooperation. Additionally, we uncovered some intriguing findings when examining the characteristics of these biomedical scholars. Contrary to our initial hypothesis that researchers without computer science background would be more likely to seek collaborators with such expertise, the data revealed the opposite pattern. Our analysis showed that among independent researchers, 16.86% possessed substantial computer science knowledge, while this proportion rose to 45.10% among those who chose to collaborate, indicating that researchers with computer knowledge were more willing to engage in interdisciplinary collaboration.

We speculate this phenomenon may stem from two factors: First, the constraints of individual collaboration networks make researchers more likely to connect with those sharing similar knowledge backgrounds. Scholars with computer expertise tend to have academic networks that include more computer scientists, facilitating such cross-disciplinary partnerships. Second, from a subjective perspective, researchers generally prefer collaborating with peers who share similar knowledge backgrounds, as this reduces communication barriers and minimizes potential misunderstandings arising from significant knowledge disparities (Lorenzetti & Rutherford, 2012). As for this, we included whether researchers themselves possess computer experience as a control variable in our regression model in Section 2.4.

### *5.2 Why AlphaFold barely contributes to fostering interdisciplinary collaboration?*

AlphaFold's limited impact primarily stems from two key reasons. First, as a specialized computational tool, AlphaFold inherently only creates collaboration demand with computer science. This specific technical attribute determines that it mainly promotes collaboration with computer scientists, while being unable to facilitate the construction of broader interdisciplinary collaboration networks. Second, with the advancement of technological democratization, the widespread adoption of various simplified tools and operational tutorials has significantly lowered the usage barriers, enabling structural biologists to independently utilize this technology for research. This consequently results in AlphaFold producing merely a 0.48% increase, demonstrating very limited effect. This phenomenon has also been verified in our event study analysis.

From the case of AlphaFold, we can see that the role of technology in promoting interdisciplinary collaboration is limited by many factors. As technology becomes more widespread and is applied more deeply, scholars gradually become able to independently master and use relevant tools to complete work that originally required interdisciplinary collaboration, thus reducing the demand for such collaboration.

Additionally, a large number of studies have shown that interdisciplinary scientific collaboration is not easy to establish or maintain (Lee et al., 2009; Liu et al., 2024). From the perspective of social identity theory, there is a clear social categorization between different disciplines (Judge & Ferris, 1993). Scholars have developed unique identities and professional norms within their own disciplines, which makes them need to invest extra time and energy to build trust in interdisciplinary collaboration (Leahey et al., 2017). Interdisciplinary collaboration requires handling issues involving at least two disciplines, which demands more effort from researchers to coordinate the requirements and resources of different disciplines (Gewin, 2014). As a result, even if scholars are motivated to use technology for interdisciplinary collaboration, the challenges of such collaboration often make them hesitate or lead them to seek collaborators with relevant technical skills only within their own discipline. Therefore, although technology provides the possibility for interdisciplinary collaboration, its role in promoting such collaboration remains very limited.

*5.3 Contributions and policy implications*

The primary contribution of this study is to provide new evidence on the role of artificial intelligence (AI) in interdisciplinary collaboration. The findings indicate that, although the application of AlphaFold significantly increased the proportion of collaborations between structural biologists and computer scientists by 0.48%, it had no significant effect on collaborations with other disciplines, which contradicts some mainstream views. Additionally, through heterogeneity analysis, the study further investigates the differences in preferences for interdisciplinary collaboration among researchers with varying levels of productivity, team sizes, and academic ages, providing new insights into the complexity of interdisciplinary collaboration.

Although mainstream views and certain examples have demonstrated that AI can promote interdisciplinary cooperation, such as medical researchers collaborating with AI experts to predict disease transmission during the COVID-19 pandemic, and the field of AI ethics fostering dialogue between technologists and scholars in the humanity and social science. However, our research indicates that the driving force of AI in promoting interdisciplinary cooperation is not as expected. Specifically, AI tools exhibit a dual nature in influencing interdisciplinary research collaboration: certain fields or studies may generate specific collaboration demands due to technical barriers, yet this dependency is simultaneously weakened by the process of technological democratization. As revealed by the classification of interdisciplinary collaboration mechanisms by Dai & Boos (2019), purely technology-based interdisciplinary cooperation will have its influence weakened for various reasons. In the future cultivation of interdisciplinary research ecosystems, there should be a greater focus on constructing deeper knowledge integration mechanisms that transcend tool dependency.

### *5.4 Limitations*

This study has several limitations. First, due to the relatively short time since the introduction of AlphaFold, the sample size available for this study is limited. This may affect the comprehensiveness and representativeness of the findings. Second, the study focuses solely on AlphaFold as a case study, which may not fully reflect the entire impact of artificial intelligence on interdisciplinary collaboration. Other AI tools and technologies may exhibit different patterns and effects in various disciplinary fields, and thus additional cases are needed to further validate the findings. Lastly, the accuracy of the disciplinary classification of researchers using the SCIBERT algorithm is 81.57%, which, although acceptable, may still have some impact on the results, especially when examining the nuances of interdisciplinary collaboration. Future research could consider employing more precise disciplinary classification methods or combining multiple classification approaches to enhance the reliability of the findings.

## 6 Conclusions

This study examines the impact of AlphaFold technology on interdisciplinary collaboration patterns among structural biologists using a combination of bibliometric analysis and causal inference methods. Based on 1,247 related papers and data from 7,700 authors from the Scopus database, we compared the treatment group (structural biologists using AlphaFold) with the control group. The results show that the application of AlphaFold increased the collaboration ratio between structural biologists and computer scientists by only 0.48%, with no significant effect on collaborations with other disciplines. Structural biologists with a background in computer science were found to have a stronger tendency for interdisciplinary collaboration. The findings reveal a dual pathway through which AI technology facilitates interdisciplinary collaboration: the technical complexity of AlphaFold drives structural biologists to seek help from computer experts, and the accumulated interdisciplinary experience through cooperation further consolidates this collaborative pattern. However, this catalytic effect is limited to collaborations between structural biology and computer science, with no significant impact on other disciplines. Heterogeneity analysis indicates that high-productivity authors, small teams, and mid-career researchers are more inclined to engage in interdisciplinary collaboration. These findings provide important insights for developing future strategies to promote interdisciplinary collaboration.

## Author contributions

Naixuan Zhao: Conceptualization, Methodology, Software, Writing – original draft.
Chunli Wei: Data curation, Methodology, Software.
Xinyan Zhang: Writing – original draft.
Jiang Li: Conceptualization, Supervision, Writing – review & editing.

## Acknowledgments

This study is financially supported by the National Social Science Fund Major Project (24&ZD321). Initial findings from this research were accepted for plenary presentation at STI 2025 (the International Conference on Science and Technology Indicators) which was hold on 3-5 September 2025.

*Linguistics*, *50*(1), 237-291.